**Amorphous ultra-wide bandgap ZnO$_x$ thin films deposited at cryogenic temperatures**

**M. Zubkins\*, J. Gabrusenoks, G. Chikvaidze, I. Aulika, J. Butikova, R. Kalendarev, L. Bikse**

*Institute of Solid State Physics, University of Latvia, Kengaraga 8, LV-1063, Riga, Latvia*

\* Corresponding author: Dr.phys. Martins Zubkins at Thin Films Laboratory, Institute of Solid State Physics, University of Latvia, Kengaraga Street 8, Riga, LV-1063, Latvia, email: zubkins@cfi.lu.lv, phone number: +37167251691, fax: +37167132778.

Crystalline wurtzite zinc oxide (*w*-ZnO) can be used as a wide band gap semiconductor for light emitting devices and for transparent or high temperature electronics. The use of amorphous zinc oxide (*a*-ZnO) can be an advantage in these applications. In this paper we report on X-ray amorphous *a*-ZnO$_x$ thin films (~500 nm) deposited at cryogenic temperatures by reactive magnetron sputtering. The substrates were cooled by a nitrogen flow through the copper substrate holder during the deposition. The films were characterized by X-ray diffraction (XRD), Raman, infrared, UV-Vis-NIR spectroscopies, and ellipsometry. The *a*-ZnO$_x$ films on glass and Ti substrates were obtained at the substrate holder temperature of approximately –100°C. New vibration bands at 201, 372, and 473 cm$^{-1}$ as well as O-H stretch and bend absorption bands in the *a*-ZnO$_x$ films were detected by FTIR spectroscopy. Raman spectra showed characteristic ZnO$_2$ peaks at 386 and 858 cm$^{-1}$ attributed to the peroxide ion O$_2^{2-}$ stretching and libration modes, respectively. In addition, the films contain neutral and ionized O$_2$ and O$_2^-$ species. The *a*-ZnO$_x$ films are highly transparent in the visible light range ($\approx$87%) and exhibit a refractive index of 1.68 at 2.25 eV (550 nm). An optical band gaps is 4.65 eV with an additional band edge absorption feature at 3.50 eV. It has been shown that the deposition on actively cooled substrates can be a suitable technique to obtain low temperature phases that cannot be deposited at room temperature.

**Keywords:** amorphous ZnO, thin films, cryogenic deposition, reactive magnetron sputtering, peroxide.

## I. INTRODUCTION

Zinc oxide (ZnO) is an important semiconductor in material science and practical applications. ZnO thin films can be deposited by the wide variety of techniques, such as magnetron sputtering, chemical vapor deposition (CVD), pulsed laser deposition (PLD), etc. ZnO crystallizes under ambient conditions in a thermodynamically stable hexagonal wurtzite-type (*w*-ZnO) structure $P6_3mc$ (No.186). However, other polymorphs, such as zinc blende, body-centered tetragonal, cubane, sodalite, and hexagonal boron nitride structures have been theoretically predicted to be stable under suitable thermodynamic conditions, and part of them have been synthesized.[1] *w*-ZnO is an extensively studied oxide with a wide direct band gap (~3.37 eV at 300 K) and has a strong tendency to naturally form as an *n*-type semiconductor.[2] It can be transformed into a highly conducting material with high visible light transmittance by proper doping.[3] All this makes *w*-ZnO thin films an important material for optoelectronic devices.

Amorphous films are of great interest because they can provide several advantages over crystalline ones, such as lower deposition temperature, higher uniformity, better compatibility with flexible substrates, and in some cases higher carrier mobility due to the absence of grain boundaries.[4] The majority of studies and applications are based on *w*-ZnO films. It is challenging to produce amorphous (*a*-) films of pure ZnO due to its rapid crystallization. Therefore, experimental structural information on *a*-ZnO is still mostly vague. Experimental studies have shown that the *a*-ZnO phase can be stabilized in relatively thin films (< 150 nm),[5,6] in alloyed Zn-Me-O films,[7,8] by chemical solution deposition,[9] or by deposition under cryogenic conditions.[10] However, some of the studies lack convincing evidence whether the films are completely amorphous or nanocrystalline. In addition to experimental investigations, Mora-Fonz *et al*.[11] calculated that a room temperature stable *a*-ZnO structure could be produced under fast cooling from melt. The expected average density of $a$-ZnO would be ~5 gcm$^{-3}$ and the coordination number of Zn and O atoms ~3.9.

Previous study[12] has shown the presence of peroxide ions $(O_2)^{2-}$ groups into the nanocrystalline and amorphous Ir doped/alloyed ZnO films. Zinc peroxide (ZnO$_2$) has a pyrite structure $Pa$-3 (No.205).[13] The structure consists of an array of ZnO$_6$ octahedra, where Zn$^{2+}$ ions are located at the central octahedron. The ZnO$_2$ decomposes into a $w$-ZnO phase (2ZnO$_2$(s) = 2ZnO(s) + O$_2$ (g)) at about 230°C at ambient pressure.[14,15] The reported fundamental band gap of ZnO$_2$ is in the range between 3.3 and 4.6 eV and indirect with the valence band maximum situated at $\Gamma$ and the conduction band minimum situated between $\Gamma$ and R.[13,16]

In this paper we study the possibility to obtain amorphous ZnO thin films deposited by reactive DC magnetron sputtering on actively cooled substrates ($< 0$°C) and investigate their composition and properties. Much emphasis is placed on the optical band gap determination of $a$-ZnO$_x$ films, since the values of the band gap differ in the literature. The films have been characterized by X-ray diffraction (XRD), Raman, FTIR, UV-Vis-NIR spectroscopies, and spectroscopic ellipsometry.

## II. EXPERIMENTAL DETAILS

$a$-ZnO$_x$ as well as $w$-ZnO thin films were deposited on soda-lime glass, Si(111), Ti, and CaF$_2$(111) substrates, by reactive DC magnetron sputtering from Zn (purity 99.99 %) target in an Ar+O$_2$ atmosphere. Several types of substrates were used to be able to study films by different characterization techniques. The substrates were ultrasonically cleaned with acetone, detergent, 2-isopropanol for 15 min each, rinsed with distilled water, and then dried under blown N$_2$ gas. The thin film deposition was performed using the vacuum PVD coater G500M (Sidrabe Vacuum, Ltd.). The coater was equipped with a substrate cooling system using a nitrogen flow through the grounded copper substrate holder. All types of substrates were stick to the holder using double-sided bonding tape. Before the deposition process, the chamber ($\approx 0.1$ m$^3$) was pumped down to base pressure below $1.3 \times 10^{-5}$ mbar by a turbo-molecular pump backed with a rotary pump. The films were deposited at three different substrate holder temperatures of +38 (without intentional heating or cooling), –42, and –103°C. The last two temperatures were obtained adjusting the nitrogen flow using the valve between the substrate holder and the liquid nitrogen container. When the necessary temperature was reached, the substrate holder was kept at this temperature for 10 min to allow all substrate temperatures to equilibrate. Then the process gases were introduced into the chamber. The Ar (purity 99.99%) and $O_2$ (purity 99.5%) gas flow rates were 10.0 sccm and 7.5 sccm, respectively. A relatively high oxygen fraction was used to stimulate amorphous structure growth. It is known from the literature that high oxygen partial pressure during the deposition of zinc oxide deteriorates crystallinity.[17,18] The pumping speed was altered by a throttle valve to set the sputtering pressure of $8.0 \times 10^{-3}$ mbar. At this pressure, process gases do not condensate on the cooled substrates, since their boiling point is well above the liquid nitrogen temperature of 77 K.[19] The target was sputtered in a constant DC mode at a power of 200 W. A planar magnetron with target dimensions 145×92×3 mm was used. The distance between the target and the substrates was approximately 9 cm. The nitrogen flow through the holder was increased slightly during the deposition to compensate the additional heating from a plasma discharge and maintain a constant temperature. After the process, the samples were kept in a vacuum until they reached room temperature without intentional heating and then removed from the chamber.

The structure of the samples was examined by an X-ray diffractometer with Cu Kα radiation, Rigaku MiniFlex 600. The film morphology was studied under a scanning electron microscope (SEM) Thermo Fisher Scientific Helios 5 UX. Raman scattering spectroscopy measurements were carried out at room temperature using a TriVista CRS Confocal Raman System. A YAG second harmonics laser (532 nm) was used as the excitation source. The incident beam power was about 4 mW. The Fourier transform infrared (FTIR) absorbance spectra were measured by using a VERTEX 80v vacuum FTIR spectrometer. The experiments were performed in the range from 100 to 6000 $cm^{-1}$, with the interferometer working in vacuum and with a resolution of 4 $cm^{-1}$. The samples on Si substrates were used for the FTIR measurements using uncoated Si as a background.

The film's transmittance and reflectance, in the range of 200 to 1200 nm, were determined by a spectrophotometer, Agilent Cary 7000. The sample was placed at an angle of 6 degrees against the incident beam, and the detector was placed at 180 degrees behind the sample to measure transmittance, and at 12 degrees in front of the sample to measure specular reflectance. Optical properties and film thicknesses were obtained by means of spectroscopic ellipsometer (SE) WOOLLAM RC2 in the spectral range from 210 to 1690 nm or from 5.9 to 0.7 eV. The main ellipsometric angles $\Psi$ and $\Delta$ were measured at the incident angles from (55-85)° with the (2-5)° step. Refractive index $n$ and extinction coefficient $k$ dispersion curves were modeled using two Gaussian and one Cody-Lorentz oscillator (CLO) functions for crystalline samples and single CLO or CLO and Tauc-Lorentz (TLO) for amorphous films.[20] Substrates without films were measured to obtain precise optical properties of the substrates, where later these data were used for the film SE data modeling. SE experimental data model-based regression analyses were performed with the WOOLLAM software CompleteEASE®.

After the film's characterization there was an indication of $ZnO_2$ phase formation in the films. To ensure a good $ZnO_2$ reference sample, nanocrystalline $ZnO_2$ powder was directly produced by a hydrothermal process using zinc acetate dehydrate as a precursor and hydrogen peroxide as an oxidizing agent. 11.4 g of zinc acetate dehydrate ($Zn(CH_3COO)_2 \cdot 2H_2O$) were dissolved in 750 ml of deionized water in a round-bottom flask with a reflux condenser and 70 ml of hydrogen peroxide ($H_2O_2$; 30 vol.%). $H_2O_2$ was added and then the solution was heated up to 100°C to conduct the hydrothermal reaction. Afterwards, the solution was cooled down to room temperature and polycrystalline $ZnO_2$ was extracted from the liquid and washed with deionized water. The material was finally dried in air for 24 hours. For the FTIR absorption measurements, a $ZnO_2$ layer on a Si substrate was precipitated from an aqueous emulsion.

## III. RESULTS AND DISCUSSION

### A. Structure and morphology (XRD and electron microscopy)

The evolution of the X-ray diffractograms, recorded over a range for 2θ of 20°–80°, as a function of deposition temperature for the films deposited on glass substrates is shown in Fig. 1. A significant diffraction maximum at around 34° and a low intensity maximum at around 71° are observed for the film deposited at room temperature (+38°C). The maximums correspond to the $w$-ZnO lattice planes (002) and (004), respectively (PDF card No.: 01-070-8072). This indicates that the film contains a crystalline phase growing preferentially with the $c$-axis perpendicular to the substrate surface. The size of crystallites in the [002] direction is 13 nm calculated from the Scherrer equation and the lattice parameter $c$ is 5.33 Å.

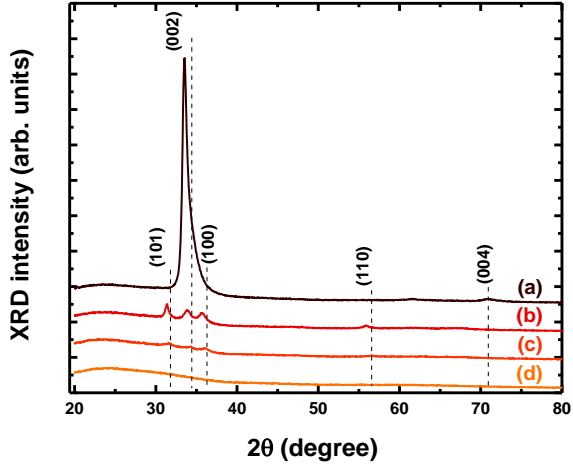

FIG. 1. X-ray diffractograms of the thin films deposited on glass as a function of deposition temperature: +38°C (a), −42°C (b), and −103°C (d). The films were deposited from a Zn target in a reactive atmosphere ($O_2$/Ar flow ratio 3/4) at pressure of $8 \times 10^{-3}$ mbar and at average power of 200 W. The figure also contains the X-ray diffractogram of the post-annealed (100°C, 2.5 h, in a vacuum) film deposited at −103°C (c). The plane indexes correspond to the wurtzite type *w*-ZnO structure according to PDF card 01-070-8072.

The films lose their crystalline phase orientation with decreasing deposition temperature, and additional maximums of the (100), (101), and (110) planes appear for the film deposited at −42°C. The crystallite size of 9 nm in the [002] and [101] directions is slightly smaller compared to the sample deposited at room temperature, but it is 22 nm in the [100] direction. The values of lattice parameters *a* and *c* are 3.30 Å and 5.29 Å, respectively. For the both samples (+38°C and −42°C) the lattice parameters are larger than the standard *w*-ZnO crystal lattice parameters *a*=3.25 Å and *c*=5.21 Å (Ref. 3) indicating tensile stress.

Further decrease of the deposition temperature degrades the crystalline structure and the films become X-ray amorphous at the deposition temperature of −103°C. The X-ray amorphous film was post-annealed in a vacuum at the temperature of 100°C for 2.5 hours to crystallize a possible $ZnO_2$ phase, but low intensity diffraction maximums (100), (002), and (101) related to $w$-ZnO structure become apparent after the annealing. The evolutions of the X-ray diffractograms with the deposition temperature for the films deposited on the Ti substrates are qualitative the same (see Fig. S1 of supplementary material). However, the films on the Si and $CaF_2$ substrates become X-ray amorphous already at –42°C (see Figs. S2 and S3 of supplementary material).

The X-ray diffractogram of the $ZnO_2$ powder (see Fig. S4 of supplementary material) does not show any impurity phases, every visible maximum can be assigned to the $ZnO_2$ phase with a $Pa$-3 (No.205) space group. Initial lattice parameters and atomic coordinates for the Rietveld refinement were taken from the paper by W. Chen $et$ $al.$[13] Rietveld refinement was done with the BGMN software[22] using Profex[23] as a graphical user interface. Rietveld refinement gave a very good fit, only some discrepancies between the calculated and experimental X-ray diffractograms can be seen in the most intense maxima of the diffractogram, but the maxima at higher angles shows an excellent fit as can be seen in the inset of Fig. S4 of supplementary material. Rietveld refinement allowed an accurate determination of the crystallite size, which was found to be 22 nm.

The films deposited at +38°C and –103°C were imaged by SEM (Fig. 2). To suppress charging, the films were coated with a 20 nm thin layer of gold to suppress charging. Densely packed grains with distinctly sharp edges were observed on the surface of the room temperature deposited film (Fig. 2(a)). The average grain size is about 50 nm. Drastic morphology change can be seen in the sample deposited at –103°C (Fig. 2(b)). The surface becomes featureless and smooth with a fine-grained structure.

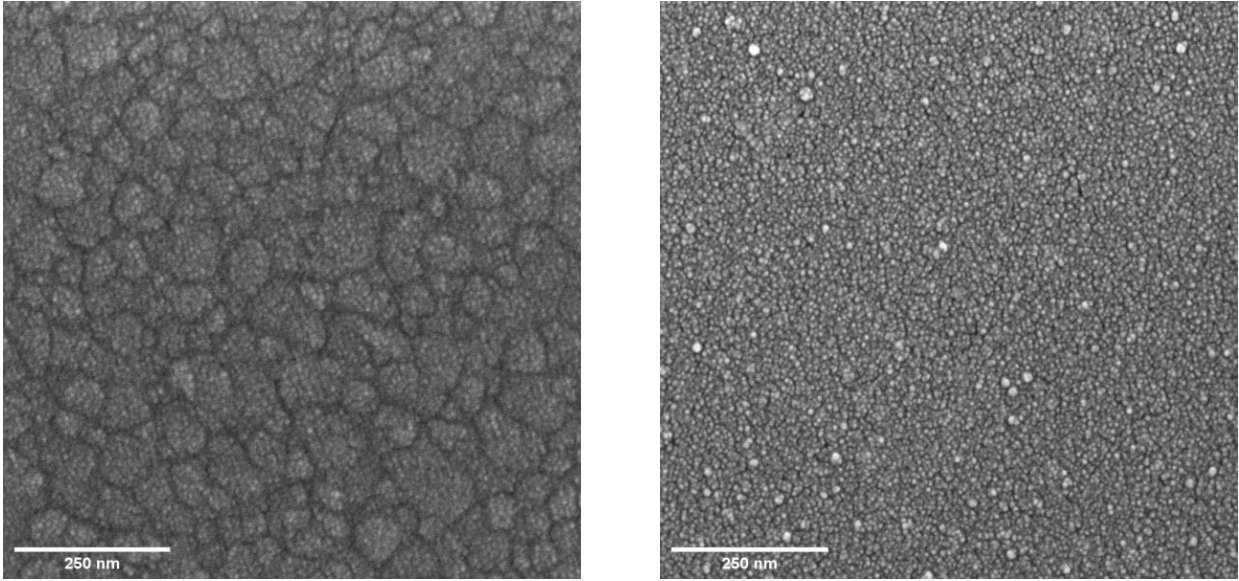

FIG. 2. SEM images of the ZnO$_x$ films deposited at two different temperatures on glass substrates: +38°C (a) and –103°C (b).

## B. Vibrational analysis (Raman and FTIR spectroscopy)

Fig. 3 shows the Raman spectra of the films deposited on Ti substrates as a function of deposition temperature. Bands at 325, 439, 569, 622, 858, 1120, and 1543 cm$^{-1}$ were observed for the crystalline films deposited at +38 and –42°C. The bands at 439 and 569 cm$^{-1}$ are assigned to the non-polar E$_2$(*high*) and polar A$_1$(LO) Raman active vibration modes, respectively, of *w*-ZnO. The band at 325 cm$^{-1}$ is a difference mode E$_2$(*high*)– E$_2$(*low*), the band at 622 cm$^{-1}$ could be tentatively assigned to A$_1$(TA+TO), and the wide band at 1120 cm$^{-1}$ originates from the second order mode 2A$_1$(LO).[24] The rest of the bands at 858 and 1543 cm$^{-1}$ cannot be explained within the framework of the bulk *w*-ZnO phonon modes.[25] These bands are caused by the presence of molecular oxygen species, and the bands may be attributed to the stretching modes of O$_2$$^{2-}$ and O$_2$, respectively.[26] The frequency of the detected band 1543 cm$^{-1}$ is lower compared to the frequency of free gaseous O$_2$ 1556 cm$^{-1}$. The immediate chemical environment affects the O–O bond. This indicates that the band does not originate from oxygen in the air. In addition, the width and intensity of the peak are greater than what can be detected from atmospheric oxygen under these conditions. The molecular oxygen incorporation into the films even at the deposition temperature of +38°C suggest incomplete dissociation of oxygen. The band at ~850 cm$^{-1}$ has been observed frequently by other groups even at higher deposition temperatures and different deposition techniques.[27]

All the Raman active bands that correspond to the *w*-ZnO phase disappear when the structure becomes X-ray amorphous, while the molecular oxygen related bands remain and even new ones appear at 386 and 1042 cm$^{-1}$. The band at 1042 cm$^{-1}$ could be assigned to the $O_2^-$ stretching, and the band at 386 cm$^{-1}$ to the $O_2^{2-}$ libration. The bands at 386 and 858 cm$^{-1}$ correspond with the Raman spectra of nanocrystalline $ZnO_2$ (Fig. 3 and Ref. 26) un suggest the formation of a $ZnO_2$ phase in the X-ray amorphous films deposited at approximately –100°C on Ti. The slight shift of the Raman peaks in the X-ray amorphous film spectrum compared to the powder spectrum can be explained by a change in the peroxide bond caused by the highly disordered structure. In principle, the stretching frequency of peroxide ions is highly dependent on the distance between the oxygen atoms and increases as the distance is reduced.

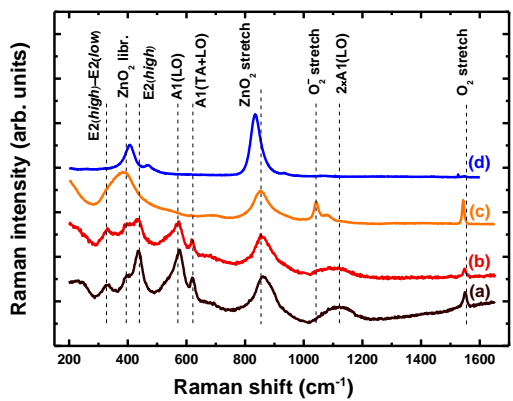

FIG. 3. Raman spectra of the ZnO$_x$ deposited on Ti as a function of deposition temperature: +38°C (a), –42°C (b), and –103°C (c). The figure also contains the Raman spectrum of the $ZnO_2$ powder (d).

During the infrared absorption measurements, the films on Si substrates were irradiated perpendicularly to the film's surface with non-polarized light. The absorption spectra of the films in the far IR region as a function of deposition temperature are shown in Fig. 4. For the room temperature (+38°C) deposited film only the transverse optical mode $E_1$(TO) of $w$-ZnO around 411 cm$^{-1}$ was observed (Fig. 4(a)) because crystallites are $c$-axis oriented. New absorption bands at 201, 372, and 473 cm$^{-1}$ (Fig. 4(b)) appear when the film structure becomes X-ray amorphous at the deposition temperature of –42°C (see Fig. S2 of supplementary material). These bands do not fit with the infrared active $w$-ZnO vibration modes. Although the band at 372 cm$^{-1}$ could be tentatively assigned to $A_1$(TO). It seems that the bands at 372 and 473 cm$^{-1}$ broaden and overlap each other, and the intensity of the band at 201 cm$^{-1}$ reduces when the deposition temperature is further decreased (Fig. 4(c)).

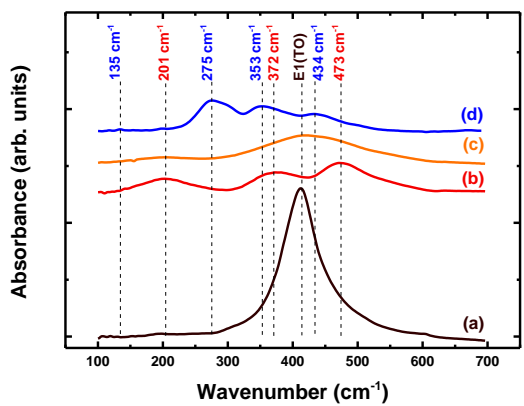

FIG. 4. Absorption spectra of the $ZnO_x$ films deposited on Si(111) in the far IR as a function of deposition temperature: +38°C (a), –42°C (b), and –103°C (c). The figure also contains the absorption spectrum of the $ZnO_2$ powder (d).

The calculations of $ZnO_2$ fundamental vibrations were performed (see supplementary material for details) to try to explain the newly observed IR bands in the films. The calculated vibration frequencies (see Table SI of supplementary material) are in good agreement with the observed Raman and IR frequencies of the produced $ZnO_2$ powder, but do not fit well with the $a$-$ZnO_x$ film's spectra, except for Raman bands that were already mentioned. If the $ZnO_2$ phase is present then the shifted bands could be explained by the amorphous structure and the presence of the O-H groups (see Fig 5).

Figure 5 shows the absorption spectra of the films in the mid IR region as a function of deposition temperature. The spectra of the films deposited on actively cooled substrates (Fig. 5(b,c)) contain the absorption bands at about 3310 and 1400 cm$^{-1}$, which are attributed to the stretching vibration of the O-H bond and the bending vibration of Zn-O-H, respectively. It is reasonable to assume that films deposited at low growth temperature contain a lot of highly reactive unfilled bonds. Most likely the O-H groups form into the films after a deposition when the air is introduced into the chamber.

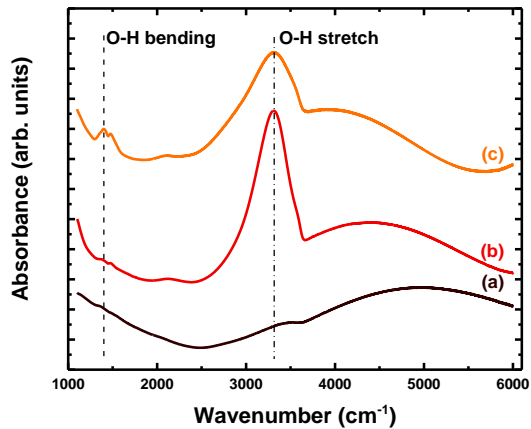

FIG. 5. Absorption spectra of the $ZnO_x$ films deposited on Si(111) in the mid IR as a function of deposition temperature: +38°C (a), –42°C (b), and –103°C (c).

## C. Optical properties

### *1. UV-Vis-IR transmittance and reflectance spectra*

Fig. 6 shows the specular transmittance and reflectance of the films deposited on soda-lime glass measured in the range from 200 to 1200 nm. All the films exhibit high transmittance in the visible light range, limited mainly by the reflectance of approximately 15% and 9% (average values in the visible range) for the crystalline and X-ray amorphous films, respectively. Thus, the average transmittance in the visible range of the amorphous film (≈87%) is higher compared to the crystalline samples (≈80%). All the spectra exhibit an interference fringe, which indicates that both the surface and the interface with a glass substrate are optically smooth. The oscillation amplitude of transmittance and reflectance curves decreases together with the structure amorphization. It is due to the lower refractive index for the X-ray amorphous film: refractive index of these films becomes comparable with the refractive index's values of the glass substrate. The sharp fall in transmittance below 400 nm is due to the onset of fundamental absorption of ZnO. The change of the film structure with the deposition temperature can be indirectly evaluated by observing a fundamental absorption edge shift. The significant blue-shift of the edge is observed when the film becomes X-ray amorphous. As soda-lime glass strongly absorbs in the UV part, the films were also deposited on the $CaF_2$ substrates to be able to determine an optical band gap by applying the Tauc model (Fig. 7). The optical band gap of ~3.43 eV for the crystalline sample is in good agreement with the bulk value of $w$-ZnO, but for the amorphous one the optical band gap is primarily determined to be ~4.97 eV. At the same time, another band edge absorption feature with an associated Tauc gap of ~3.22 eV is observed. This point will be discussed in more detail together with the ellipsometry data analysis. It is known from other reports that the band gap of $ZnO_2$ is in the range from 3.3 to 4.6 eV.[13,16]

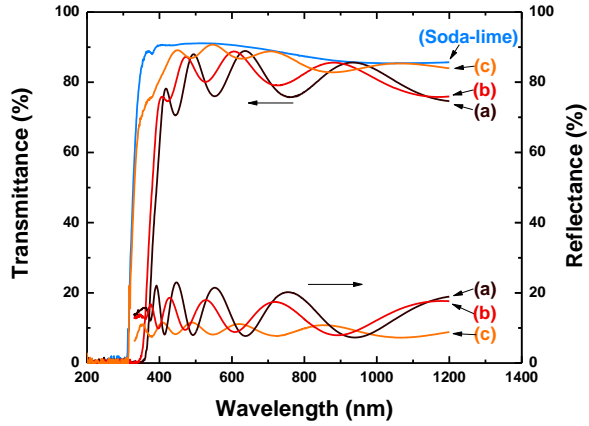

FIG. 6. Transmittance and reflectance of the ZnO$_x$ films on glass substrates in the range of 200 to 1200 nm as a function of deposition temperature: +38°C (a), –42°C (b), and –103°C (c). The figure also contains the transmittance of soda-lime glass.

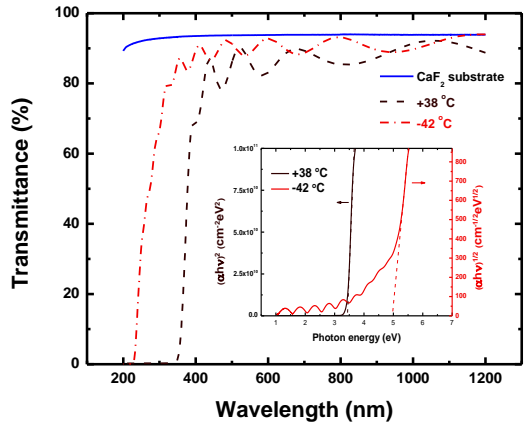

FIG. 7. Transmittance of the ZnO$_x$ films on CaF$_2$(111) in the range of 200 to 1200 nm as a function of deposition temperature. The figure also contains the transmittance of CaF$_2$. Inset shows a Tauc plot for the crystalline and X-ray amorphous films.

## 2. Spectroscopic ellipsometry

An example of the model compared to experimental SE data is present in Fig. S5 of supplementary material. The lowest MSE (mean square error) was observed for the crystalline sample deposited at +38°C (MSE = 2.1). The best fit was found by applying an optical gradient model (optical constants profile within the film) with the optical constant inhomogeneity at 1.5%. It is a small variation of optical constants, what can be explained with the strong XRD (002) maximum and very weak (004) maximum for this sample suggesting that it has low physical properties variation due to the strong crystallographic orientation. An ideal model (substrate/ideal film/surface roughness) fit gave the best result with the MSE of approximately 6 for the films deposited at –42°C. The MSE was reduced from 9 to 7 for the films deposited at –103°C by introducing in the model an interface layer between the substrate and the film: the thickness of the interface layer estimated by SE is approximately 15 nm. For other films the interface layer was not giving any improvement in the model. Some of parameters obtained from SE are summarized in the Table I.

Table I. The roughness, interphase, film thickness, optical band gap energy $E_g$, Urbach energy $E_u$ and refractive index $n$ at 1.960 eV (632.8 nm) obtained for the ZnO$_x$ films deposited at different temperatures using SE data.

| Deposition temperature (°C) | Roughness (nm) | Interphase (nm) | Thickness (nm) | $E_g$ (eV) | $E_u$ (meV) | n at 1.96 eV (632.6 nm) |
|---|---|---|---|---|---|---|
| +38 | 3.72 ± 0.01 | – | 492.19 ± 0.03 | 3.117 ± 0.002 | 77 ± 2 | 1.951 ± 0.002 |
| −42 | 5.73 ± 0.03 | – | 473.11 ± 0.11 | 3.191 ± 0.022 | 951 ± 7 | 1.923 ± 0.004 |
| −103 | 3.26 ± 0.05 | 15.5 ± 0.5 | 655.40 ± 0.32 | 3.50 ± 0.15 | 1100 ± 50 | 1.666 ± 0.006 |
| | | | | 4.65 ± 0.40 | | |

Refractive index $n$ and extinction coefficient $k$ dispersion curves for the films deposited at three different temperatures on glass substrates are given in Fig 8. There is a significant change in the refractive index and optical band gap when the deposition temperature decreases from −42 to −103°C, which is related with the phase change from crystalline to X-ray amorphous. The refractive index at 1.96 eV decreases from 1.92 to 1.67 (see Table I). The refractive index values from 1.66 to 1.71 in the visible range of the X-ray amorphous film is in good agreement with $ZnO_2$ data from the literature.[28,29]

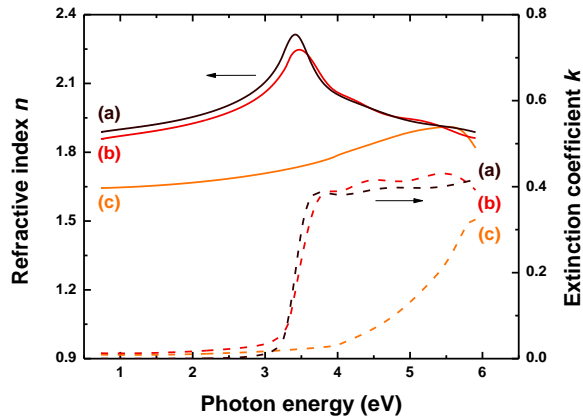

FIG. 8. Refractive index $n$ and extinction coefficient $k$ dispersion curves for the $ZnO_x$ films deposited at three different temperatures on glass substrates: +38°C (a), −42°C (b), and −103°C (c).

The optical band gap energy $E_g$ was obtained directly from the CLO and TLO as $E_g$ is one of the oscillator parameters. The CLO includes Urbach absorption term ($E_u$ - Urbach energy) to model absorption below $E_g$. The choice of CLO was because all samples represented significant Urbach type absorption below $E_g$ due to the presence of a free exciton just below the gap. The optical dispersion curves for the sample deposited at –103°C was generated using two oscillator model (CLO and TLO) according to the transmission spectra (Fig. 6(c)), where slight absorption "shoulder" around 350 nm is observed. Other samples were fitted using single CLO and two Gaussian oscillators. The $E_g$ values for the three films obtained from CLO and TLO are summarized in the Table I.

The two band gap values for the films deposited at –103°C are representing the absorption edges at 3.50 eV and 4.65 eV. The obtained band gap at 4.65 ± 0.40 eV is in good agreement with the data obtained from absorption spectra (Fig. 7) fitted using Tauc relation. Similar two $E_g$ values can be obtained using Tauc relation to fit extinction coefficient curve simulated with CLO+TLO. The presence of two $E_g$ values might be related with two phases present in the material: 1) $a$-ZnO$_2$ with the band-gap of about 4 eV,[13,16] and 2) $a$-ZnO having a gap of 3.4 eV.[6] More detailed investigation should be done to clarify this two-band gap nature of $a$-ZnO$_x$ films.

The $E_g$ values are lower compared with the transmission spectra (Fig. 7) results for the films deposited at +38°C and –42°C (Tab. I). It should be taken into account that it is difficult to determine the fundamental optical band gap for the material with a significant Urbach tail even in case of CLO application. It is more correct to call $E_g$ the near band gap value rather than the fundamental optical band gap for the materials with Urback tails. The fundamental optical band gap extraction from extinction coefficient $k$ dispersion curves can be done more correctly considering three different fits to the $k$ curves[30] since there are three different contributions for the absorption coefficient of the dielectric thin film: 1) the Urbach tail, which dominates the absorption in the region just below the exciton peak, 2) a free exciton peak, which appears strong just below the energy gap, and 3) the charge transfer excitations from the valence band to the conduction band, which dominates the high energy region. Considering just said, in cases of steep extinction coefficient increase at around of $E_g$ value and small increase at energies above $E_g$ (like for the films deposited at +38°C and –42°C, see Fig. 8(a,b)), the fundamental band gap values are very near to oscillator peak position energy $E_o$ value rather to the $E_g$ values obtained directly from CLO or TLO.[31] The peak position energy values for the films deposited at +38°C and –42°C are 3.49 eV and 3.66 eV, respectively. These values are similar to the $E_g$ values obtained from the absorption spectra.

There is also a slight blueshift from 3.11 to 3.19 eV in the crystalline films as the deposition temperature is reduced from +38°C to –42°C. A similar blueshift phenomenon of optical band gap was also observed in ZnO thin films deposited on sapphire substrate and on fused quartz by MOCVD,[32] and in our previous studies on ZnO powders.[33] The main factor contributing to the absorption edge broadening in crystalline materials is the exciton–phonon coupling dynamic disorder. Imperfections and disordering, especially pronounced in amorphous materials, bring additional broadening due to static disorders.[34] $E_u$ reflects the localized state distribution in the valence band tail. As crystallinity decrease with decrease of the film deposition temperature, imperfections and disordering increase and thus the increase of $E_u$ is observed (Table I). The $E_u$ values obtained for the crystalline $w$-ZnO$_x$ thin films deposited at +38°C are comparable with the data found in literature.[31] High $E_u$ values for the films deposited at low temperatures can be explained with fact that these films have significant absorption even at low photon energies (< 3 eV). It can be explained with presence of imperfections and defects in these films.

## IV. CONCLUSIONS

This paper investigates a challenging subject of amorphous ZnO deposition and its structure as well as optical properties. The crystalline ($w$-) and X-ray amorphous ($a$-) ZnO$_x$ thin film growth at cryogenic temperatures by reactive DC magnetron sputtering was demonstrated.

The X-ray amorphous structure of ZnO$_x$ can be obtained by reducing the deposition temperature, but the exact temperature at which the amorphization takes place depends on the substrate material. The presence of the ZnO$_2$ phase was observed by Raman and infrared spectroscopies. The observed vibration bands are wide and shifted compared to the ZnO$_2$ powder spectra, which is explained by the highly disorder structure and the additional defects as O-H groups and various oxygen species.

The $a$-ZnO$_x$ films have considerably lower refractive index and higher optical band gap compared to $w$-ZnO. The optical band gaps were determined by a Tauc plot and an ellipsometry data analysis. The methods gave good band gap values match; they confirmed the significant blueshift with the amorphization. Two $E_g$ values were obtained for $a$-ZnO$_x$ thin film suggesting that this material is composed of two phases: more detailed investigation should be done to clarify this band gap nature. The possibility to grow amorphous films at lower deposition temperature offers different advantages like compatibility with flexible substrates for technologies where high temperatures are not allowed.

**ACKNOWLEDGMENTS**

Financial support provided by Scientific Research Project for Students and Young Researchers Nr. SJZ/2018/8 realized at the Institute of Solid State Physics, University of Latvia is greatly acknowledged.